\documentclass[pdflatex,sn-mathphys-num]{sn-jnl}

\usepackage{makecell}
\usepackage{graphicx}%
\usepackage{multirow}%
\usepackage{amsmath,amssymb,amsfonts}%
\usepackage{amsthm}%
\usepackage{mathrsfs}%
\usepackage[title]{appendix}%
\usepackage{xcolor}%
\usepackage{textcomp}%
\usepackage{manyfoot}%
\usepackage{booktabs}%
\usepackage{algorithm}%
\usepackage{algorithmicx}%
\usepackage{algpseudocode}%
\usepackage{listings}%

\theoremstyle{thmstyleone}%
\theoremstyle{thmstyletwo}%

\theoremstyle{thmstylethree}%

\begin{document}

\title[]{Review on recent progress in the study of the $N^*(920)$ subthreshold singularity and the $\sigma/f_0(500)$ meson}


\author[1]{\fnm{Qu-Zhi} \sur{Li}}\email{liquzhi@scu.edu.cn}

\author[1]{\fnm{Zhiguang} \sur{Xiao}}\email{xiaozg@scu.edu.cn}

\author[1]{\fnm{Han-Qing} \sur{Zheng}}\email{zhenghq@scu.edu.cn}

\affil[1]{\orgdiv{Institute for Particle and Nuclear Physics, College of Physics}, \orgname{Sichuan University}, \orgaddress{
\city{Chengdu}, \postcode{610065}, \state{Sichuan}, \country{P.R. China}}}
%
%


\abstract{We summarize recent results on studies of $\pi\pi$ and $\pi N$ scatterings. They include the finding of a negative-parity nucleon pole with a mass lower than the nucleon mass, and the pole trajectory of $f_0(500)$ as the pion mass varies. The results are obtained from model-independent dispersion analyses. We also study the thermal properties of $f_0(500)$ based on the $O(N)$ $\sigma$ model and $N/D$ method.}




\maketitle

\section{Introduction}\label{sec1}
In recent years, considerable progress has been made in understanding low-energy QCD, through the study of $\pi\pi$ and $\pi N$ scatterings. In particular, a novel subthreshold pole has been found in the $S_{11}$ channel, with the pole position, $\sqrt{s}=(918\,\pm\, 3)\,-\,i\,(163\,\pm\, 9)\,$MeV. This result is obtained by a Steiner or Roy--Steiner equation analysis~\cite{Cao:2022zhn}, as inspired by an earlier analysis using the product representation for partial wave amplitudes~\cite{Wang:2017agd}. This result is totally unexpected, since it sounds completely counterintuitive to imagine a resonant pole below the `lowest' lying and hence stable nucleon pole.   

The very existence of the $N^*(920)$ pole poses a significant challenge to our understanding of non-perturbative QCD in the $\pi N$ sector. We will discuss such an issue in Sections 2 and 4, though there are  still a lot left to be understood on this pole.

In addition to  the $\pi N$ sector, there are also new results on the $f_0(500)$/$\sigma$ resonance.
There has been a long history in the study of this particle. Due to page limitation here we only refer to earlier review papers,  
 such as Refs.~\cite{Pelaez:2015qba,Yao:2020bxx,Pelaez:2021dak} for previous developments in this area. In a series of recent publications~\cite{Cao:2023ntr,Lyu:2024lzr,Lyu:2024elz}, we have achieved a better understanding of the property of this very important particle, since it is widely believed that $f_0(500)$ plays a crucial role in the spontaneous chiral symmetry breaking mechanism.  We will summarize our new results on $f_0(500)$
in Section 3. Section 4 is dedicated to the conclusions and perspectives for future explorations, especially the thermal property of $N^*(920)$. 

\section{The \texorpdfstring{$N^*$(920)}{N*(920)} pole}\label{sec2}

\begin{figure}
    \centering
    \includegraphics[width=0.9\linewidth]{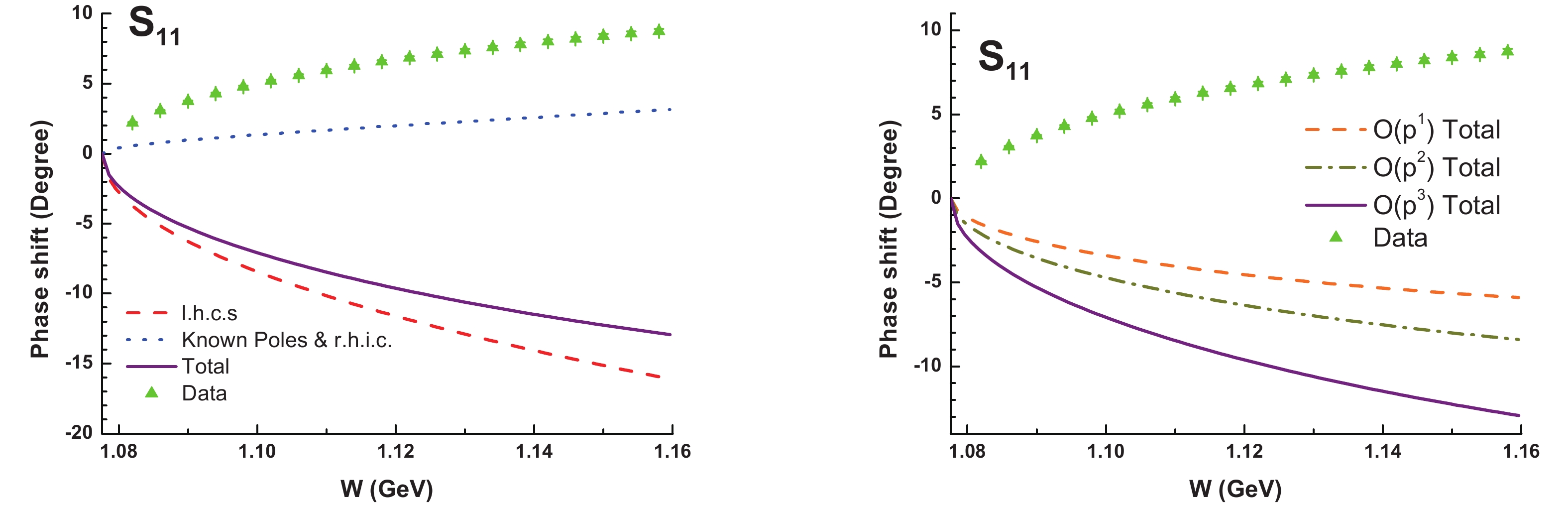}
    \caption{The different contributions to the $S_{11}$ phase shift~\cite{Wang:2017agd}.}
    \label{fig:S11-Wang}
\end{figure}

The existence of $N^*(920)$ pole was originally suggested
in Ref.~\cite{Wang:2017agd}, where the authors analyzed all six low partial waves (two $s$ and four $p$ waves) for $\pi N$ scattering amplitudes. 
According to the product representation~\cite{Xiao:2000kx,He:2002ut,Zheng:2003rw,Zhou:2006wm},  different contributions to the $\pi N$ scattering phase shifts are additive. Particularly, 
resonance pole and virtual state pole contributions are positive whereas bound state pole contribution and `left-hand cut' contribution are negative. The left-hand cut or the background contribution is known in potential scattering theory to be $\delta_{b.g.}=-kR$~\cite{Hu:1948zz}, where $k$ is the channel momentum and $R$ the interaction range of the potential (exact for truncated potentials~\cite{Regge:1958ft}). In relativistic scattering theory the background contribution is represented by an integral on non-unitary cuts. It is estimated using tree-level $\pi N$ chiral amplitudes in Ref.~\cite{Wang:2017agd}, and $O(p^3)$ amplitudes in Ref.~\cite{Wang:2018nwi}, both of which are qualitatively in agreement with the potential scattering case.
After all known contributions to the total phase shift are carefully taken into account,
it is revealed that a large
discrepancy with the experiment persists in
the $S_{11}$ channel among all six partial waves~\footnote{
Actually, a significant mismatch also arises in the $P_{11}$ channel. This is compensated by an accompanying virtual pole contribution, in 
addition to the nucleon bound state pole. The existence of such a virtual pole is mandated by the `elementariness' of the 
nucleon~\cite{Zhang:2009bv,Cao:2020gul}.
} This discrepancy can only be accounted for by the existence of
second-sheet poles. As such, a resonance pole located at { 
$\sqrt{s}=895\,(81)\,-\,i\, 164\,(23)\,$ MeV } is needed to accommodate the experimental data. In addition, it is
examined that an estimation of the left-hand cut contribution at $O(p^3)$ level only slightly
alters the pole position~\cite{Wang:2017agd,Wang:2018nwi}. See Fig.\ref{fig:S11-Wang} for an illustration. Furthermore, the $N^*$ pole
has been subsequently confirmed in the $K$-matrix~\cite{Ma:2020sym} and $N/D$ method~\cite{CPC:10} unitarization approximation  of the $\pi N$
scattering amplitude. 

Actually, there are even earlier analyses in the $S_{11}$ channel of $\pi N$ scatterings, which suggest the possible existence of a pole with a mass around 1100 MeV (and a width around 400 MeV) which is very near to the $\pi N$ threshold rather than below the nucleon mass~\cite{Azimov:1970ei,Garcia-Recio:2003ejq,Doring:2009uc}.
Nevertheless, it is worth pointing out
that a $K$-matrix unitarization approximation suffers some fundamental-level
problems. There are three fundamental properties of the $S$ matrix: unitarity,
analyticity, and crossing symmetry. The $K$-matrix unitarization often violates
the latter two~\cite{Qin:2002hk,Guo:2007ff,Guo:2007hm} (for a pedagogical
review, see also Ref.~\cite{Yao:2020bxx}). Hence, the $K$-matrix unitarization
method and its variations are generally not suitable for the discussion of
distant poles.  Similar problems also exist in the other unitarization approaches such as Bethe-Salpeter equation, unitarized $\chi$PT. As a result, though a similar near threshold pole was found  
in~\cite{Doring:2009uc}, the authors failed to recognize the pole as a genuine state but tended to interpret it as an effect that mimics the missing cut~\cite{Doring:2009yv,Doring:2009uc}. A comment on Ref.~\cite{Doring:2009uc} can be found in Ref.~\cite{CPC:10}.

After the discovery of the subthreshold resonance pole, its physical properties have also been investigated. {In Ref.~\cite{Ma:2020hpe} its coupling to $N\gamma$ and $N\pi$ systems is studied in the $\gamma N\to \pi N$ process. It is found that the $N^*N\gamma$ coupling is similar to that of $N^*(1535)N\gamma$, and the $N^*N\pi$ coupling is significantly greater than the $N^*(1535)N\pi$ coupling. The latter is reasonable since it is well known that $N^*(1535)$ couples very weakly to the $N\pi$ system. }

However, accepting such a subthreshold pole seems to be difficult primarily for two reasons.  First, given a history of over six decades of the study on the $\pi N$ scattering, it is hard to expect that there could be any undiscovered physics left.  Indeed, it is truly remarkable --- and somewhat puzzling --- that it was not until very recently that the subthreshold singularity structure of $\pi N$  scattering amplitudes was settled down~\cite{Cao:2022zhn,Li:2021oou} (see also \cite{Doring:2025sgb} for a recent review). 
Second, the stable nucleon is well known as the lowest-lying baryonic state. Thus, the existence of any resonance pole below the $\pi N$ threshold is counterintuitive, if not in contradiction to physical rules. That is, if a state has no states to decay into, why does it develop a width as the $N^*(920)$ behaves? The answer to the second question is that the usual intuitive understanding of `width' is only a concept in \emph{perturbation} theory. 
A formal, non-perturbative, and mathematically rigorous definition of `width' is only related to the imaginary part of the $S$ matrix pole position in the $k$-plane and does not necessarily relate to the concept of decay phase space.\footnote{Only in weak coupling limit one finds  the `width' relating to the decay phase space.} In a rigorous quantum theory, a state with a complex pole mass simply means nothing more than a generalized Hamiltonian eigenstate with a non-normalizable wave function.  

\begin{figure}
    \centering
    \includegraphics[width=0.9\linewidth]{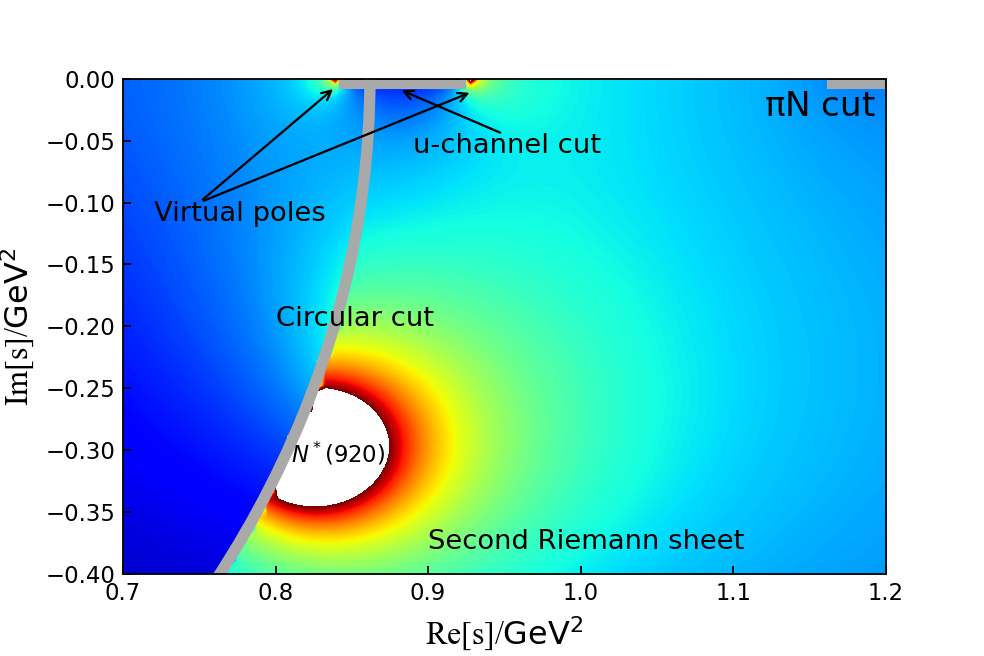}
    \caption{The pole position of  $N^*(920)$ obtained by solving Roy-Steiner equation.~\cite{Cao:2022zhn}}
    \label{fig:N*920}
\end{figure}

 Steiner or Roy-Steiner equations grounded on axiomatic quantum field theory are most rigorous and powerful tools in investigating low-energy hadron scatterings non-perturbatively. These equations  inherently incorporate unitarity, analyticity and crossing symmetry properties of the $S$ matrix in a set of integral equations. By solving these integral equations numerically with the experimental low energy and high energy input, the $N^*(920)$ pole is unambiguously identified on the second Riemann sheet at $\sqrt s=(918\,\pm\, 3)\,-\,i\,(163\,\pm\, 9)\,$ MeV as shown in Fig.~\ref{fig:N*920}, which provides the most powerful proof for the existence of this resonance pole~\cite{Cao:2022zhn}. This result, later confirmed by Ref.~\cite{Hoferichter:2023mgy}, firmly established the existence of the $N^*(920)$ pole.

\begin{figure}
     \centering
     \includegraphics[width=0.7\textwidth]{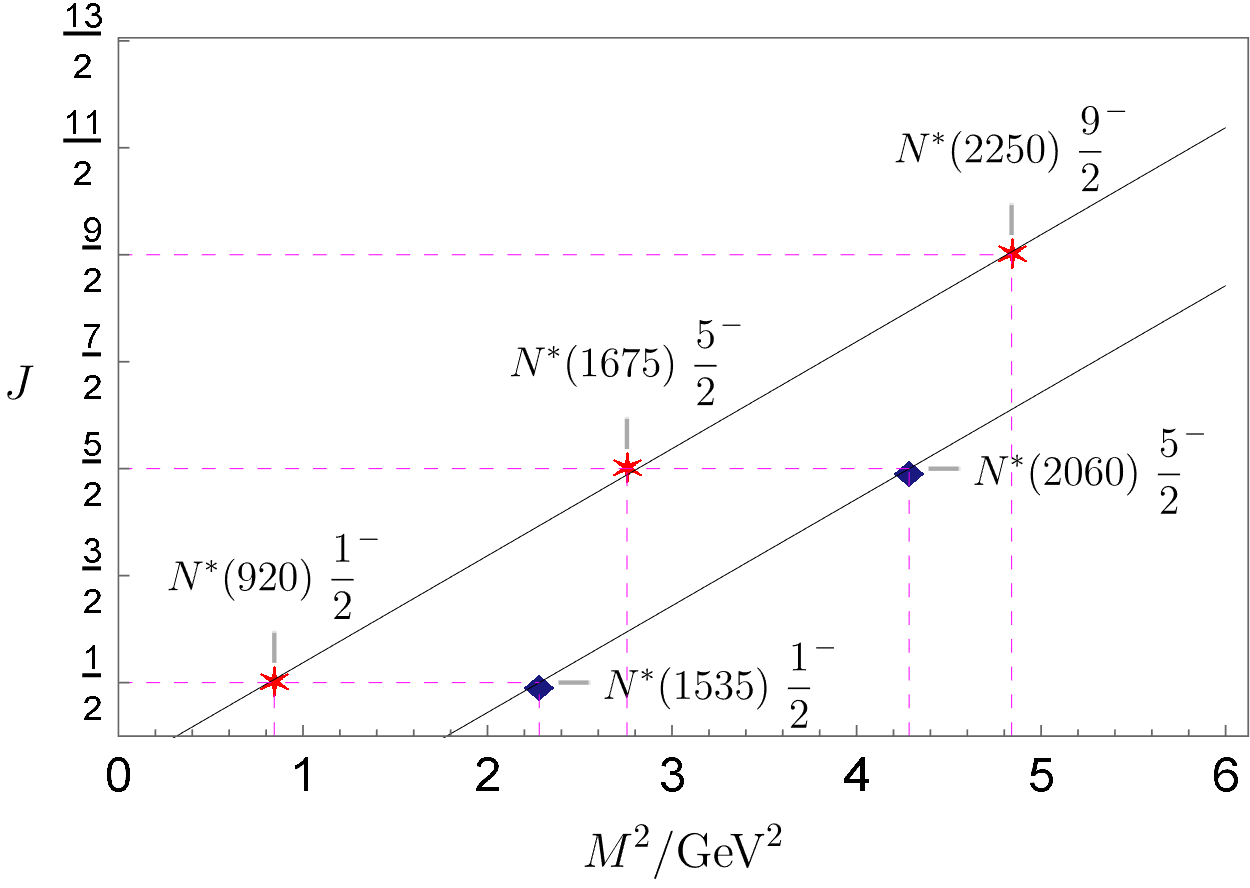}
     \caption{Nucleon Regge trajectories.}
     \label{fig1}
 \end{figure}

\begin{figure}
    \centering
    \includegraphics[width=0.5\linewidth]{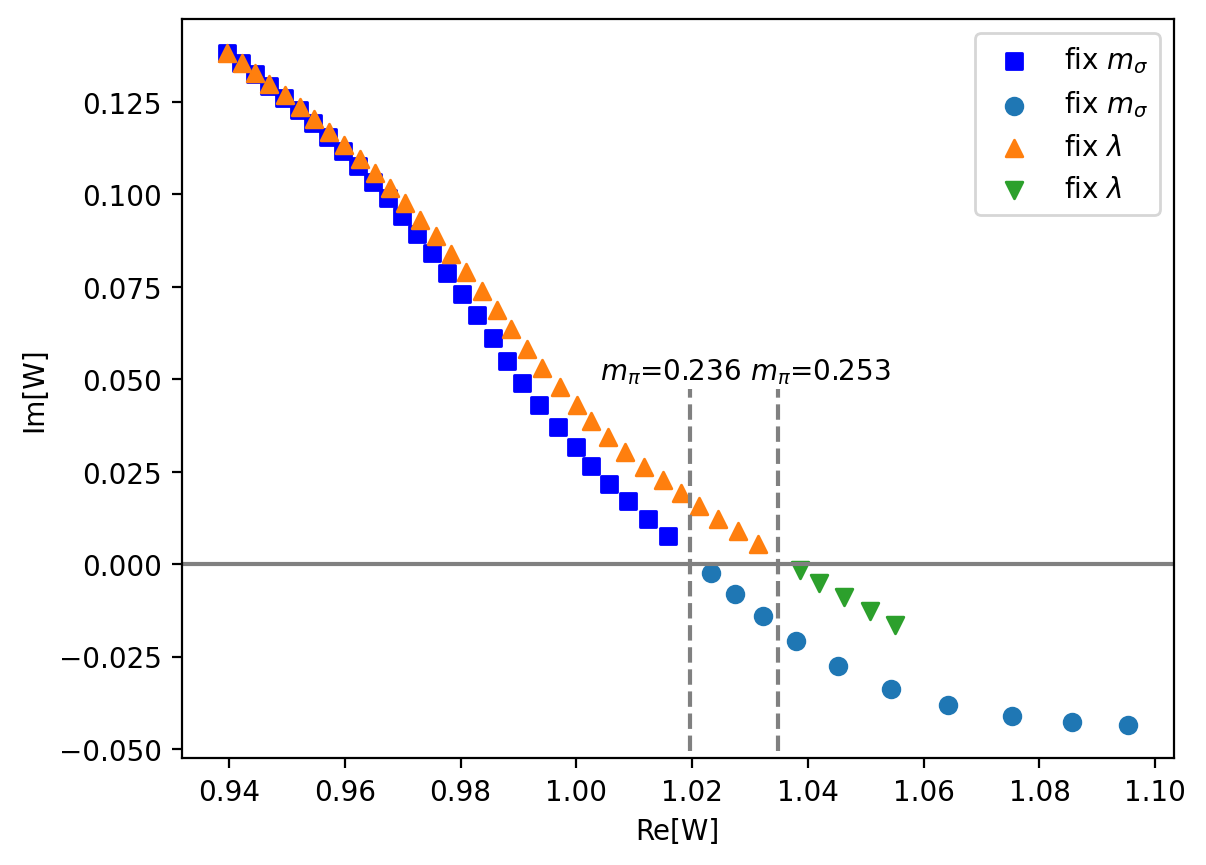}
    \caption{$N^*(920)$ pole trajectories on the $W\equiv \sqrt s$ plane at tree level with $m_\pi$ ranging from $0.138\mathrm{GeV}$  to $0.270\mathrm{GeV}$ for two different scenarios. The solid circles and inverse triangles below the real axis represent the $N(920)$ pole trajectories that move through the $u$-cut onto the unphysical sheet defined by the logarithmic branch points of the this cut.  See Ref.~\cite{Li:2025fvg} for details.}
    \label{fig:N920-mpi}
\end{figure}

 Nevertheless, much work remains to be done in understanding the physical
 properties of such a pole hidden far away from the physical region. Before
 proceeding, it is worth pointing out that the nucleon Regge trajectory as
 inspired by the generalized McDowell symmetry~\cite{Collins:1977jy}  appears to
 necessitate such a subthreshold pole, as can be seen in Fig.~\ref{fig1}.
 Another way to probe the nature of $N^*(920)$ is to study its pole trajectory
 as the pion mass varies, which could be useful for future
 lattice studies. In~\cite{Li:2025fvg}, using the linear $\sigma$-model and the
 $N/D$ method as the unitarization scheme, at tree level, it is found that the $N^*(920)$ pole
 moves towards the $u$-cut on the real axis in the $\sqrt s$-plane as pion mass
 increases (see Fig.\ref{fig:N920-mpi} for an illustration). It
 finally disappears from the second sheet and moves across the $u$-cut at a certain point to the
 other sheet defined by the logarithmic branch points.
  This may not necessarily be the final destiny of the pole, however.
 In principle, when the pole approaches the real axis, there is no
compelling reason it has to hit the $u$-cut. For example, it might be possible that
 the $N(920)$ poles fall at some point between the $u$-cut and the 
 right-hand cut on the real axis and then split into two virtual states.  
 Further one-loop $N/D$ calculation is necessary and still in progress.
 

More interestingly, it would be very desirable to investigate the role $N^*(920)$ plays at high temperatures. This topic relates to the important topic of how chiral symmetry is restored at high temperatures, i.e., is there a parity doublet or not and will $N^*(920)$ play a role in it? We postpone the discussion to the end of the next section.

\section{Dynamical properties of $\sigma$ resonance in the case of unphysical pion masses and high temperatures}

The existence of $f_0(500)/\sigma$ resonance was once a topic of debate for decades, until it was finally settled using the dispersion techniques~\cite{Zhou:2004ms,Caprini:2005zr,Garcia-Martin:2011nna}~\footnote{For a review, one is referred to, for example, Ref.~\cite{Yao:2020bxx}.}. 
Nevertheless, to the best of the authors' knowledge, there has never been conclusive evidence that the $f_0(500)$ resonance takes the pivotal role of the prominent $\sigma$ meson in spontaneous chiral symmetry breaking, though it may already be rather widely embraced~\footnote{See for example, Refs.~\cite{Guo:2006br}}. 

\begin{figure}[h]
\centering
\includegraphics[width=0.5\textwidth,angle=-0]{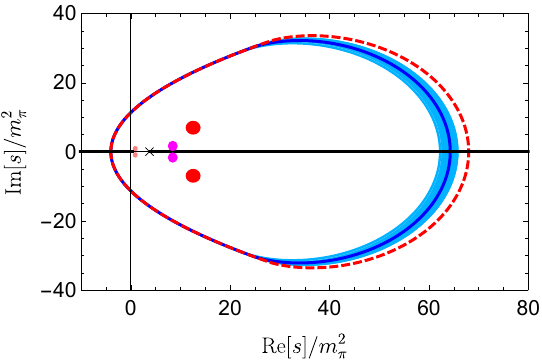} 
\caption{Validity domain of extended Roy equation for $m_\pi=391$~MeV. 
The dashed red boundary represents the validity domain by dropping the effects of the bound state $\sigma$, and the blue boundary corresponds to the complete validity domain within uncertainty from the location of the $\sigma$. 
The $\sigma$ pole now becomes a bound state pole represented by a cross. Discussions on  other poles can be found in Ref.~\cite{Cao:2023ntr}.} \label{fig.roydomain}

\end{figure}

\begin{figure}[h]
\centering
\includegraphics[width=0.5\textwidth,angle=-0]{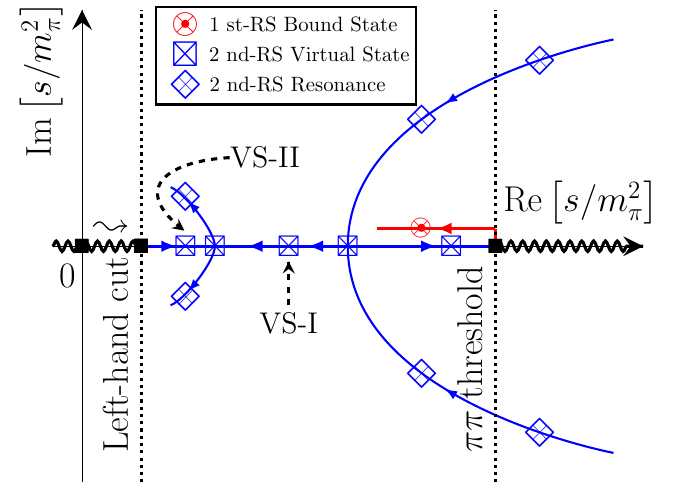} 
\caption{The sugguested trajectory of the $\sigma$ pole on the second Riemann sheet of the $s$ plane with varying $m_\pi$ in \cite{Cao:2023ntr}. \label{fig.trajectory}}
\end{figure}

Recently, considerable progress has been made in this direction. In lattice QCD
studies, more and more phase-shift data have been accumulated with different
input pion masses. An investigation on the $f_0(500)$ pole trajectory with
respect to varying pion masses becomes available. 
In
\cite{Briceno:2016mjc,Briceno:2017qmb}, $f(500)$ pole positions as extracted from lattice data
using a simple $K$-matrix method, are found to be a resonance pole at around $\sqrt
s \sim (487 \sim 809)-i(136 \sim 304)$ MeV for $m_\pi=239$MeV, and a bound state
pole at around $\sqrt s \sim 758\pm 4 $MeV for $m_\pi=391$MeV. 
There are also some theoretical work using unitarization method such as IAM and the unitarized $\chi$PT to discuss the resonance pole trajectories in $\pi\pi$ scattering~\cite{Hanhart:2008mx,Pelaez:2010fj,Hanhart:2014ssa}. The result 
shows that as the pion mass increases, the $f(500)$
pole moves toward the real axis and become two virtual poles. Then one virtual 
pole moves towards the threshold and across the threshold becoming a bound state, which qualitatively agrees with the lattice result.
However, as
previously pointed out, widely used $K$ matrix unitarizations often have severe
shortcomings, which do not match the lattice data that are obtained from ab
initio calculations.  In addition, since crossing symmetry is widely believed to
be important in determining the $f_0(500)$ resonance, incorporating crossing
symmetry in the analysis of lattice data is necessary. A first trial in this direction is in~\cite{Gao:2022dln}
where the PKU representation was utilized in determining the $f_0(500)$ pole
position from the lattice data, in which the left-hand cut contribution was
estimated using the $O(p^4)$ $\chi$PT $\pi\pi$ scattering results and the 
crossing symmetry was imposed by the BNR relations.
The result shows that  a virtual state is accompanying the bound state when
$m_\pi=391$MeV. However, since $f_0(500)$ becomes a bound state, it will also
generate a left-hand cut through the cross-channel bound-state exchange processes, which 
does not present
in the  estimation~\cite{Gao:2022tlh}. These
unsatisfactory situations are resolved in~\cite{Cao:2023ntr}, where modified Roy
equations are constructed to accommodate the new situation of $\sigma$ bound
states for large pion masses, and the phase shifts obtained by numerically 
solving these equations agree well with the lattice data. It is found that at 
$m_\pi=239$MeV, the $f_0(500)$ pole moves to $\sqrt s\sim 543-i250$MeV and
at $m_\pi=391$MeV, the $f_0(500)$ pole becomes a bound state at $\sqrt s\sim 759^{+7}
_{-16}$MeV.
Besides the bound state found when $m_\pi=391$MeV, there is no virtual pole but a subthreshold resonance pole located at $\sqrt s= 269^{+40}_{-25} - i211^{+26}_{-23}$MeV. 
All the singularities on the second sheet are inside the Lehmann ellipse as shown in Fig.~\ref{fig.roydomain}.
To understand this subthreshold resonance pole, it is argued that another right-moving virtual state (VS-II) appears from the branch point of the left-hand cut generated by the $f_0(500)$ after it becomes a bound state when pion mass is large. 
The subthreshold resonance pole is generated after VS-II meets the left-moving virtual state pole (VS-I) originated from $f_0(500)$ poles after they hits the real axis below the threshold when pion mass becomes large enough. The suggested pole trajectory is shown in Fig.~\ref{fig.trajectory}.

More recent results by HadSpec collaboration using the $K$-matrix
method in analyzing the lattice data shows that 
at $m_\pi \sim 330$ MeV, $\sigma$ already becomes a shallow bound state,
whereas at $m_\pi \sim 283$ MeV it may become a virtual state or a
subthreshold resonance depending on the parameterization~\cite{Rodas:2023gma}. 
A further lattice study using Roy-equation analysis at $m_\pi \sim 239$ and $283$ MeV was also done in Ref.~\cite{Rodas:2024qhn}. It is found that for the former, the lowest $f_0$ state remains a resonance as before, whereas in the latter case the state was claimed, though not definitively, to become a virtual state at around $\sqrt s\sim (522\sim 562)$MeV, accompanied by a ``noisy" pole close to the left-hand cut on the second Riemann sheet.  


The above results can be viewed as directly coming from QCD, though the
$f_0(500)$ pole trajectory alone does not tell much about the resonance
properties. Inspired by the results of Ref.~\cite{Cao:2023ntr},  the linear $O(N)$ $\sigma$ model is
re-investigated in great detail in
Refs.~\cite{Lyu:2024elz,Lyu:2024lzr}. The leading $1/N$ order partial wave 
amplitude respects unitarity but does not have left-hand cut. To incorporate
both crossing symmetry and unitarity, $N/D$ method is used which recovers the
leading $1/N$ order partial wave amplitude but also includes the cross-channel
contribution which is of next to leading $1/N$ order. It is found that the
$\sigma$-pole trajectory is satisfactorily reproduced, at qualitative level,
with all the desired features of low energy QCD singularity structure, including
the subthreshold resonance pole discussed previously.  A little modification to
the picture proposed in~\cite{Cao:2022zhn} is that the VS-II pole could appear
from the left-hand cut before the $\sigma$ pole turns into a bound state. The lesson one learns about in this study is that
$O(N)$ linear $\sigma$ model does represent low energy QCD in
$IJ=00$ channel satisfactorily, and hence one intends to conclude that
$f_0(500)$ plays the role of the $\sigma$ particle.~\footnote{Of course, in
practice the situation has to be more complicated, such as the inclusion of
vector mesons, and other high order terms into the effective lagrangian, etc..
But these extra terms should not alter the
picture that a linear realization of
chiral symmetry is realized in nature. }

\begin{figure}
    \centering
    \includegraphics[width=0.4\textwidth]{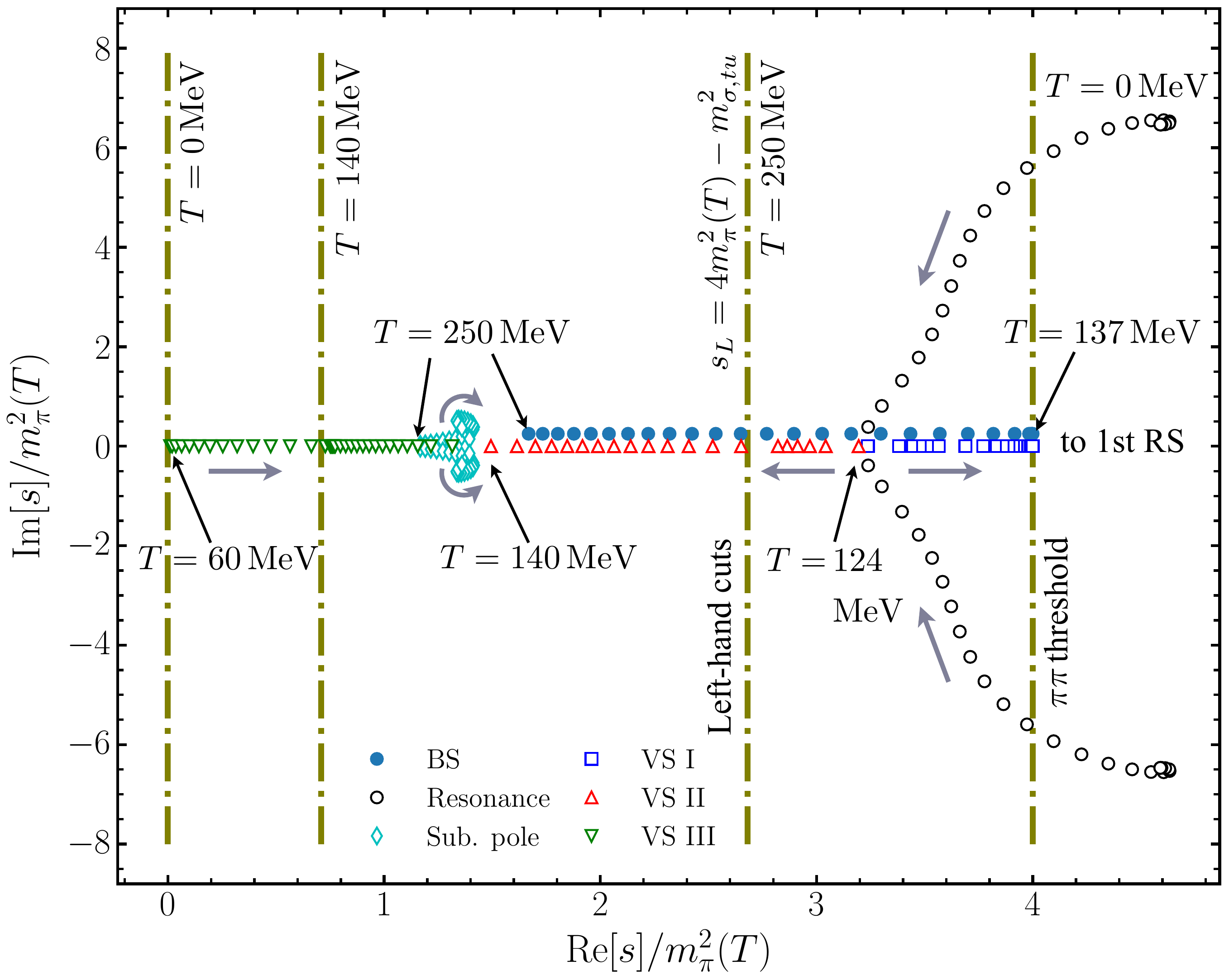}

    \caption{The $\sigma$ pole  trajectory obtained in $N/D$ modified $O(N)$ model with finite temperature~\cite{Lyu:2024elz}. When the temperature increases, $\sigma$ turns into two virtual states (VS I\&II) and then VS I becomes a bound state (BS) when temperature is above about $137$MeV. The third virtual state pole (VS III) generated from the branch point of the left-hand cut will meet VS II on the real axis, and then they become a pair of subthreshold poles.} 
    \label{fig:finiteT_ND_traj}   
\end{figure}

The success of the $O(N)$ linear $\sigma$ model
certainly deserves more respects and attentions. 
A further investigation is to study the thermal properties of the $\sigma$ pole (namely the $f_0(500)$ resonance) in the $O(N)$ model. 
Similar to the zero-temperature case, the cross-channel contribution to the amplitude would also affect the behavior of the poles in the finite temperature scattering amplitude. The $N/D$ method can be extended to the finite temperature case~\cite{Lyu:2024elz}. It is rather surprising to have found that thermal trajectory of the $\sigma$ pole is qualitatively very similar to the trajectory with varying $m_\pi$ as shown in Fig.~\ref{fig:finiteT_ND_traj}.  For example, when $T\simeq 137$MeV
$\sigma$ turns into a bound state from a virtual pole as well.~\footnote{Similar result is also obtained in \cite{Patkos:2002xb,Patkos:2002vr} using the leading order result of the $O(N)$ model, but since the crossed channels are not included,  there are no subthreshold poles generated from the branch point of the left-hand cut in these analyses.} Also the thermal vacuum property is  carefully investigated and it is found that in linear $\sigma$ model the vacuum still remains stable at the moment when $\sigma$ turns into a bound state~\cite{Lyu:2024elz,Lyu:2024lzr}.  It is noticed that the $\sigma$ pole not only turns into a bound state, but will also eventually approach the pion-pole position. This actually is not an accident, but dictated by the asymptotically restored  $O(N)$ symmetry ($SU(2)\times SU(2)\sim O(4)$ in reality) at the high-temperature limit. 

It is worth noticing that such a restoration is hard to be fulfilled in usual thermal $\chi$PT calculations, since $\chi$PT is constructed just in the broken phase. Nevertheless, there are also some hints of the chiral restoration from $\chi$PT calculations. The chiral condensate is calculated to be decreasing with temperature~\cite{Gasser:1986vb}, and  the scalar and pseudoscalar susceptibility are estimated to be degenerate around a critical temperature within the unitarized $\chi$PT~\cite{GomezNicola:2013pgq,Ferreres-Sole:2018djq}. However, though the $\sigma$ pole trajectory obtained using the unitarized $\chi$PT shows a decreasing $\sigma$ mass with increasing temperature and seems to display a trend towards the real axis at temperatures roughly above 150MeV~\cite{Fernandez-Fraile:2007hkz,Cabrera:2008tja,Gao:2019idb}~\footnote{To produce this behavior the Adler zero must be incorporated into the IAM formalism, otherwise the behavior is quite  different~\cite{Dobado:2002xf}.}, near the chiral restoration temperature, the $\sigma$ pole would not turn to a bound state or tend to the pion mass, which demonstrates the limitation of unitary approximation based on $\chi$PT.

The realization of chiral symmetry becomes much subtler when nucleons are involved. The restoration of chiral symmetry at high temperatures, in the chiral limit, would require parity doubling of nucleons and one may wonder the role of $N^*(920)$ may play here. The subtlety is that, if the nucleon became massless in the chiral symmetric phase, there is no need to introduce a new parity partner for the nucleon, because a massless fermion can be decomposed into two helicity mass eigenstates with opposite parity (see for example \cite{Weinberg:1996kr}).  On the other hand, a nucleon does not need to be massless in the chiral symmetric phase. This can be achieved at the effective lagrangian level by explicitly introducing the parity partner of the nucleon~\cite{BWLee:1972,Detar:1988kn,Jido:2001nt, Chen:2022zgm}. 
The chiral partner of the nucleon is usually expected to be $N^*(1535)$ (see for example Ref.~\cite{Gallas:2009qp,Kong:2024xia}).  In lattice studies of thermal QCD, there is evidence that the $N^*(1535)$ becomes degenerate with the nucleon at high temperatures~\cite{Aarts:2015mma,Aarts:2017rrl}. What role the $N^*(920)$ pole may play in this process would become an interesting question.  

{The lattice $\pi K$ scattering data are also examined within Roy-Steiner equations at $m_\pi=391\mathrm{MeV}$
recently~\cite{Cao:2024zuy}. There are two points worth mentioning. First,
similar to the $\sigma$,  $K^*(892)$ becomes a bound state at
$m_\pi=391\mathrm{MeV}$, leading to a modification of the Roy-Steiner equations.
Second, contrary to the $K$-matrix prediction that the $\kappa$ becomes a
virtual state~\cite{Dudek_2014}, the result of the Roy equation analysis shows that the
$\kappa$ remains a broad resonance. }

\section{Conclusions and Outlooks}

In this mini-review we have discussed recent progresses being made concerning low-energy chiral physics of QCD, including the new understanding of the $f_0(500)$ particle with varying pion masses and temperatures. Maybe more interestingly, the newly established low-lying negative parity nucleon pole raises more questions concerning its properties, and especially the possible role it may play in the realization of chiral symmetry.  

\bmhead{Acknowledgements}

The authors would like to thank Xiong-Hui Cao for valuable discussions.  They are also grateful for the support from China National Natural Science Foundation
under contract Nos. 12335002, 12375078. This work is  also supported in part by ``The Fundamental Research Funds for the Central Universities".



%



\bibliography{sn-bibliography}

\end{document}